\documentclass[runningheads]{llncs}
\usepackage{graphicx}
\usepackage{amsmath,amssymb,amsfonts}
\usepackage{graphicx}
\usepackage{textcomp}
\usepackage{algorithm}
\usepackage[bb=boondox]{mathalfa}
\usepackage[noend]{algpseudocode}
\usepackage{color}
\usepackage{multirow}
\usepackage[font=small,labelfont=bf]{caption}

\usepackage{tikz}

\usepackage[caption=false]{subfig}

\usepackage{tabularx, booktabs}
\newcolumntype{L}[1]{>{\raggedright\arraybackslash}p{#1}}
\newcolumntype{C}[1]{>{\centering\arraybackslash}p{#1}}
\newcolumntype{R}[1]{>{\raggedleft\arraybackslash}p{#1}}
\PassOptionsToPackage{unicode}{hyperref}
\PassOptionsToPackage{naturalnames}{hyperref}

\newcommand{\wb}{\mathit{WB}}
\renewcommand{\sb}{\mathit{SB}}
\newcommand{\bas}{\mathit{bas}}

\newcommand{\source}{{A_s}}
\newcommand{\target}{{A_t}}
\newcommand{\pa}{\Delta}
\newcommand{\pc}{{\rho}}
\newcommand{\one}{{\mathbb{1}}}
\newcommand{\zero}{{\mathbb{0}}}

\newcommand{\hd}{\sf{hd}}
\renewcommand{\ts}{\mathit{TS}}
\newcommand{\update}{{\xi}}
\newcommand{\naturals}{\mathbb{N}}
\newcommand{\path}{{\sigma}}

\newcommand{\post}{{\sf post}}
\newcommand{\reach}{{\sf reach}}

\newcommand{\flag}{\mathit{isValid}}

\newcommand{\True}{\mathit{True}}
\newcommand{\False}{\mathit{False}}

\usetikzlibrary{automata,shapes.multipart}
\usetikzlibrary[arrows,decorations.pathmorphing,backgrounds,positioning,fit,petri]
\usetikzlibrary{fit}
\tikzset{
  bend angle=30,
  ->,
  shorten >=1pt,
  node distance=1.3cm and 1.3cm,
  on grid,
  auto,
  initial where=above,
  initial text=,
  initial distance=0.6cm,
  inner sep=0.5mm,
  smallstate/.style={circle,draw},
  rootstate/.style={rectangle,draw},
  openstate/.style={},
  succstate/.style={font={$\surd$}},
  bendanglelarge/.style={bend angle=45},
  bendanglesmall/.style={bend angle=10},
  dim/.style={lightgray},
  diag/.style={red},
  bisim/.style={red,dashed,-,bend left},
  loop above/.style={in=110,out=70,loop,distance=0.5cm},
  loop left/.style={in=200,out=160,loop,swap,distance=0.5cm,looseness=1},
  loop right/.style={in=20,out=-20,loop,distance=0.5cm,looseness=1},
  loop below/.style={in=290,out=250,loop,distance=0.5cm},
  attr/.style={draw,fill=black!15,rounded corners}
}
\begin{document}
\title{Sequential Control of Boolean Networks with Temporary and Permanent Perturbations}
\titlerunning{Sequential Temporary and Permanent Control of Boolean Networks}
\author{Cui Su\inst{1} \and
Jun Pang\inst{2,1} }
\authorrunning{C. Su and J. Pang}
%
\institute{Interdisciplinary Centre for Security, Reliability and Trust, University of Luxembourg, Esch-sur-Alzette, Luxembourg \\ 
\and
Faculty of Science, Technology and Medicine, University of Luxembourg, Esch-sur-Alzette, Luxembourg\\
\email{firstname.lastname@uni.lu}}
\maketitle              
\begin{abstract}
Direct cell reprogramming makes it feasible to reprogram abundant somatic cells into desired cells. 
It has great potential for regenerative medicine and tissue engineering.
In this work, we study the control of biological networks, modelled as Boolean networks, 
to identify control paths driving the dynamics of the network from a source attractor (undesired cells) to the target attractor (desired cells). 
Instead of achieving control in one step, 
we develop attractor-based sequential temporary and permanent control methods (AST and ASP) to identify a sequence of interventions
that can alter the dynamics in a stepwise manner. 
To improve their feasibility, both AST and ASP only use biologically observable attractors as intermediates. 
They can find the shortest sequential paths and guarantee $100\%$ reachability of the target attractor.  
We apply the two methods to several real-life biological networks and compare their performance with the attractor-based sequential instantaneous control (ASI). 
The results demonstrate that AST and ASP have the ability to identify a richer set of control paths with fewer perturbations than ASI,
which will greatly facilitate practical applications.

\end{abstract}

\section{Introduction}
\label{sec:introduction}
Direct cell reprogramming, also called transdifferentiation, has provided a great opportunity for treating the most devastating diseases that are caused by a deficiency or defect of certain cells. 
It allows us to harness abundant somatic cells and transform them into desired cells to restore the structure and functions of damaged organs. 
However, the identification of efficacious intervention targets hinders the practical application of direct cell reprogramming.

Conventional experimental approaches are usually prohibited due to the high complexity of biological systems and the high cost of biological experiments~\cite{L16}. 
Mathematical modelling of biological systems paves the way to study mechanisms of biological processes and identify therapeutic targets with formal reasoning and tools. 
Among various modelling frameworks, 
Boolean network (BN) has a distinct advantage~\cite{KS69,Kauffman69a}. 
It provides a qualitative description of biological systems and thus evades the parametrisation problem,
which often occurs in quantitative modelling, such as networks of ordinary differential equations (ODEs).
In BNs, molecular species (genes, transcription factors, etc.) are assigned binary-valued nodes, 
being either `0' or `1'. 
The value of `0' describes the absence or inactivate state of a specie, 
whereas `1' represents the presence or activate state. 
Activation/inhibition regulations between species are encoded as Boolean functions, 
which determine the evolution of the nodes. 
The dynamics of a BN evolves in discrete time steps under one of the updating schemes, such as {\it synchronous} or {\it asynchronous} updating schemes.
The asynchronous updating scheme is considered more realistic than the synchronous one,   
since it randomly updates one node at each time step and therefore can capture different biological processes at different time scales~\cite{PHPS05}. 
The long-run behaviour of the network dynamics is described as {\it attractors}, 
to one of which the network eventually settles down. 
Attractors are used to characterise cellular phenotypes or functional cellular states~\cite{HS01}, 
such as proliferation, differentiation or apoptosis etc.~\cite{HS01}.
In the context of BNs, direct cell reprogramming is equivalent to a {\it source-target control} problem: 
identifying a set of nodes, the perturbation of which can drive the network dynamics from a source attractor to the desired attractor.

The non-determinism of the asynchronous dynamics of BNs contributes to a better depiction of biological systems. 
As a result, it makes the control problem more challenging and renders the control methods designed for synchronous BNs inapplicable~\cite{KSK13,ZKF13}. 
Another major obstacle to the control of BNs is the infamous state explosion problem --- the state space is exponential in the size of the network. 
It prohibits the scalability and minimality of the control methods for asynchronous BNs~\cite{ZA15,MHP16}. 
The limitations of the existing methods motivate us to work on efficient and efficacy methods for the minimal source-target control of asynchronous BNs. 
There are different strategies to solve the control problem. 
Based on the control steps, we have {\it one-step control} and {\it sequential control}. 
One-step control applies all the perturbations simultaneously for one time, 
while sequential control identifies a sequence of perturbations that are applied at different time steps. 
In particular, we are interested in the sequential control that only adopts attractors as intermediates, 
called {\it attractor-based sequential control}. 
Rapid development of gene editing techniques enables us to realise the control with different perturbations, including {\it instantaneous, temporary and permanent perturbations}. 
So far, we have developed methods for the minimal one-step instantaneous control (OI)~\cite{PSPM18,PSPM19},
the minimal one-step temporary and permanent control (OT and OP)~\cite{SPP19b}, and the attractor-based sequential instantaneous control (ASI)~\cite{MSHPP19}. 
In this work, we focus on the attractor-based sequential temporary and permanent control methods (AST and ASP).

Due to the intrinsic diversity and complexity of biological systems, no single control method can perfectly suit all cases. 
Thus, it is of great importance to explore more strategies to provide a number of cautiously selected candidates for later clinical validations. 
AST and ASP integrate promising factors: attractor-based sequential control 
and temporary/permanent control. 
Attractor-based sequential control is more practical than the general sequential control~\cite{MSPPHP19}, where any state can play the role of intermediate states. 
Moreover, temporary and permanent controls have proved their potential in reducing the number of perturbations~\cite{SPP19b}. 
In this work, we continue to develop efficient methods to solve the AST and ASP control problems. 
We have applied our methods to several biological networks to show their ability in finding new control paths with fewer perturbations compared to our previous methods~\cite{PSPM18,PSPM19,SPP19b,MSHPP19}.  
We believe our new methods can provide a better understanding of the mechanism-of-action of interventions and improve the efficiency of translating identified reprogramming paths into practical applications.

\section{Preliminaries}
\label{sec:preliminaries}

In this section, we give preliminary notions of Boolean networks. 
\subsection{Boolean networks}
\label{ssec:bn}
A Boolean network (BN) describes elements of a dynamical system with binary-valued nodes and interactions between elements with Boolean functions. It is formally defined as:
\begin{definition}[Boolean networks]
A Boolean network is a tuple $G = (X,F)$ where $X=\{x_1,x_2,\ldots, x_n\}$,
such that $x_i, i\in \{1,2, \ldots, n\}$ is a Boolean variable and $F=\{f_1,f_2,\ldots,f_n\}$ is a set of Boolean functions over $X$.
\end{definition}

For the rest of the exposition, we assume that an arbitrary but fixed network $G=(X,F)$ of $n$ variables is given to us.  
For all occurrences of $x_i$ and $f_i$, we assume $x_i$ and $f_i$ are elements of $X$ and $F$, respectively.
A {\em state} $s$ of $ G$ is an element in $\{0,1\}^n$.
Let $ S$ be the set of states of $ G$. 
For any state $s=(s[1],s[2],\ldots,s[n])$, and for every $i\in \{1,2, \ldots, n\}$, the value of $s[i]$,
represents the value that the variable $x_i$ takes when the network is in state $s$.
For some $i\in \{1,2, \ldots, n\}$, suppose $f_i$ depends on $x_{i_1},x_{i_2},\ldots, x_{i_k}$.
Then $f_i(s)$ denotes the value $f_i(s[i_1],s[i_2],\ldots, s[i_k])$. 
For two states $s,s'\in S$, the {\em Hamming distance} between $s$ and $s'$ is denoted as $\hd(s,s')$. 
\begin{definition}[Control] \label{def:control}
A control $C$ is a tuple $(\zero,\one)$, where $\zero, \one \subseteq [n]$ and $\zero$ and $\one$ are mutually disjoint (possibly empty) sets of indices of nodes of a Boolean network $G$. 
The size of the control $C$ is defined as $|C|=|\zero|+|\one|$. 
Give a state $s \in S$, the application of $C$ to $s$ is defined as a state $s'=C(s)$ ($s'\in S$), 
such that $\zero=\{i\in \{1,2, \ldots, n\} \mid s'[i]=0=1-s[i]\}$ and $\one=\{i\in \{1,2, \ldots, n\} \mid s'[i]=1=1-s[i]\}$.
\end{definition}
\begin{definition}[Boolean networks under control] \label{def:BNcontrol}
Let $C=(\zero,\one)$ be a control and $G=(X,F)$ be a Boolean network. 
The Boolean network $G$ under control $C$, denoted as $G|_C$, is defined as a tuple $G|_C=(\hat{X},\hat{F})$, 
where $\hat{X}=\{\hat{x}_1,\hat{x}_2,\ldots, \hat{x}_n\}$ and $\hat{F}=\{\hat{f}_1,\hat{f}_2,\ldots,\hat{f}_n\}$, 
such that for all $i \in \{1,2, \ldots, n\}$: \\
(1) $\hat{x}_i=0$ if $i\in \zero$, $\hat{x}_i=1$ if $i\in \one$, and $\hat{x}_i=x_i$ otherwise; \\
(2) $\hat{f}_i=0$ if $i\in \zero$, $\hat{f}_i=1$ if $i\in \one$, and $\hat{f}_i=f_i$ otherwise.
\end{definition}

The state space of $G|_C$, denoted $ S|_C$ is derived by fixing the values of the variables in the set $C$ to their respective
values and is defined as $ S|_C=\{ s\in S\ |\  s[i]=1 \text{ if } i\in \one \text{ and }  s[j]=0 \text{ if } j\in \zero\}$. Note that $ S|_C\subseteq  S$.
For any subset $S'$ of $S$ we let $ S'|_C = S'\cap S|_C$. 

\subsection{Dynamics of Boolean networks}
\label{ssec:dynamics}
In this section, we define several notions that can be interpreted on both $G$ and $G|_C$. 
We use the generic notion $G=(X,F)$ to represent either $G=(X,F)$ or $G|_C=(\hat{X},\hat{F})$. 
A Boolean network $G=(X,F)$ evolves in discrete time steps from an initial state $s_0$.
Its state changes in every time step according to the update functions $F$ and the update scheme.
Different updating schemes lead to different dynamics of the network~\cite{MPSY18,ZH14}.
In this work, we are interested primarily in the {\em asynchronous updating scheme} 
-- at each time step, one node is randomly selected to update its value based on its Boolean function. 
We define asynchronous dynamics formally as follows:
\begin{definition}[Asynchronous dynamics of Boolean networks]\label{def:dynamics}
Suppose $s_0\in S$ is an initial state of $G$.
The asynchronous evolution of $G$ is a function $\update: \naturals \rightarrow \wp( S)$
such that $\update(0)=\{s_0\}$ and for every $j\geq 0$,
if $s\in\update(j)$ then $s'\in \update(j+1)$ is a possible {\em next state} of $s$
iff either $\hd(s,s') = 1$ and $s'[i]=f_i(s)=1-s[i]$  
or $\hd(s,s')=0$ and there exists $i$ such that $s'[i]=f_i(s)=s[i]$.
\end{definition}

It is worth noting that the asynchronous dynamics is non-deterministic and 
thus it can capture biological processes happening at different classes of time scales.  
Henceforth, when we talk about the dynamics of $G$, we shall mean the asynchronous dynamics as defined above.
The dynamics of a Boolean network can be described as a {\em transition system (TS)}.
\begin{definition}[Transition system of Boolean networks]\label{def:ts}
The transition system of a Boolean network $G$, denoted as $\ts$, is a tuple $(S,E)$,
where the vertices are the set of states $ S$ and for any two states $s$ and $s'$
there is a directed edge from $s$ to $s'$, denoted $s \rightarrow s'$
iff $s'$ is a possible next state of $s$ according to the asynchronous evolution function $\update$ of $ G$.
\end{definition}

A {\em path} $\path$ from a state $ s$ to a state $ s'$ is a (possibly empty) sequence of transitions from $ s$ to $ s'$.
Thus, $\path =  s_0\rightarrow  s_1\rightarrow\ldots\rightarrow  s_k$, where $ s_0= s$ and $ s_k= s'$.
A path from a state $ s$ to a subset $ S'$ of $ S$ is a path from $ s$ to any state $ s'\in  S'$.
For a state $ s\in S$, $\reach(s)$ denotes the set of states $ s'$ such that there is a path 
from $s$ to $s'$ in $\ts$ and can be defined as the fixpoint of the successor operation 
which is often denoted as $\post^*$. Thus, $\reach(s)=\post^*(s)$.

The long-run behaviour of the dynamics of a Boolean network is characterised as {\it attractors}, defined as follows.

\begin{definition}[Attractor]\label{def:attractor}
An attractor $A$ of $\ts$ is a minimal non-empty subset of states of $ S$ such that for every $ s\in A, \reach(s)=A$.
\end{definition}

Any state which is not part of an attractor is a transient state.
An attractor $A$ of $\ts$ is said to be reachable from a state $ s$ if $\reach(s)\cap A\neq\emptyset$.
The network starting at any initial state $s_0 \in S$ will eventually end up in one of the attractors of $\ts$ and remain there forever unless perturbed. 
Thus, attractors are used to hypothesise cellular phenotypes or cell fates. 
We can easily observe that any attractor of $\ts$ is a bottom strongly connected component of $\ts$. 

For an attractor $A$, 
we define {\it the weak basin and the strong basin} of $A$ to imply the commitment of states to $A$.
\begin{definition}[Weak basin and strong basin]\label{def:basins}
The weak basin of~$A$ is defined as $\bas_\ts^W(A) = \{s\in S\ |\ \reach(s)\cap A\neq \emptyset\}$; 
and the strong basin of~$A$ is defined as $\bas_\ts^S(A) = \{s\in S\ |\ \reach(s)\cap A\neq \emptyset~\text{and}~\reach(s)\cap A' = \emptyset, A' \neq A  \}$.  
\end{definition}
Intuitively, the weak basin of $A$, $\bas_\ts^W(A)$, includes all the states $s$ from which there exists at least one path to $A$. 
It is possible that there also exist paths from $s$ to other attractor $A'~(A' \neq A)$ of $\ts$,  
while the notion of strong basin does not allow this. 
The strong basin of $A$, $\bas_\ts^S(A)$, consists of all the states from which there only exist paths to $A$. 

\begin{figure}[!t]
\centering
\begin{minipage}[b]{0.45\linewidth}
\centering
\centering
\includegraphics[width=.95\textwidth]{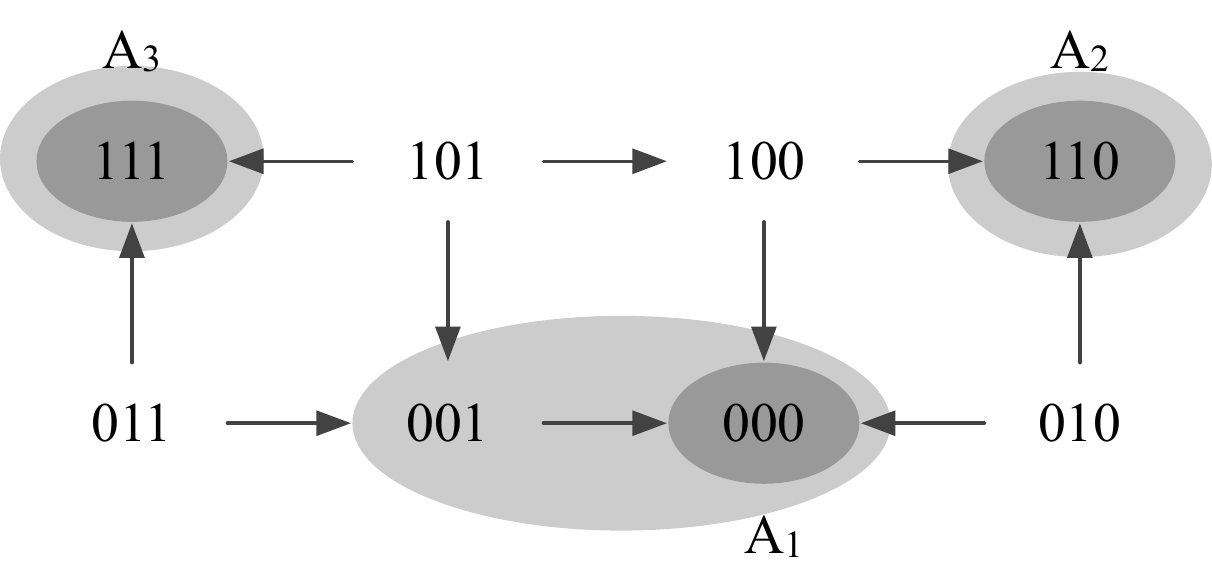}
\\(a) 
\end{minipage}
\begin{minipage}[b]{0.45\linewidth}
\centering
\includegraphics[width=.95\textwidth]{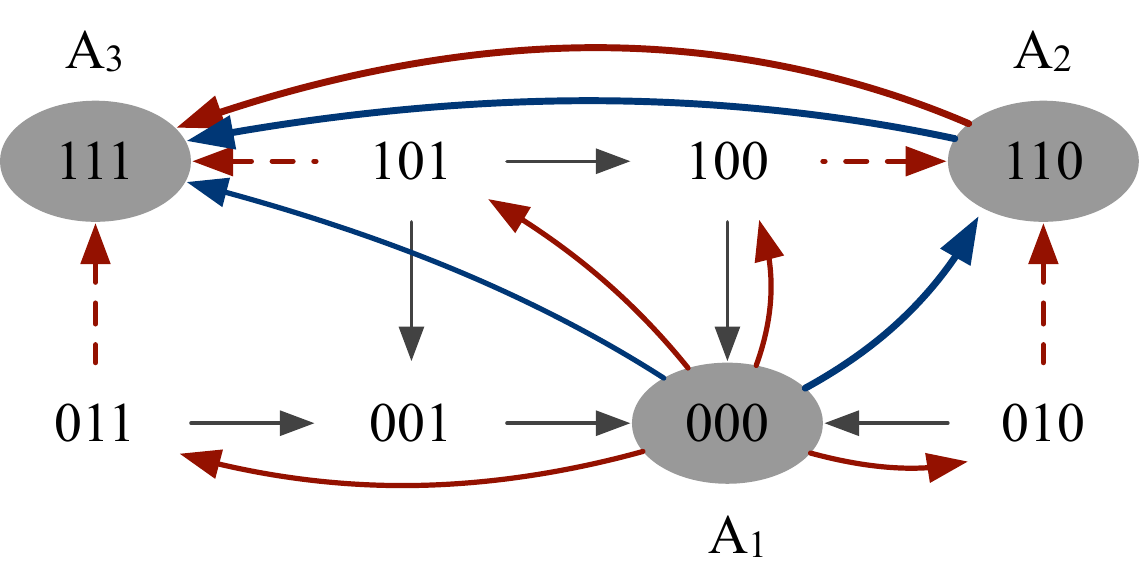}
\\(b) 
\end{minipage}
\caption{(a) The transition system of the Boolean network of Example~\ref{eg:bn}; and (b) the control paths of Example~\ref{eg:bn2}. The blue and red arrows represent the control with instantaneous and temporary/permanent perturbations, respectively. }
\label{fig:example}

\end{figure}

\begin{example} \label{eg:bn}
Consider a network $G=(X,F)$, where $X=\{x_1,x_2,x_3\}$, $F=\{f_1,f_2,f_3\}$,
and $f_1=x_2$,
$f_2=x_1$ and
$f_3=x_2 \land x_3$. 
The transition system $\ts$ is given in Fig.~\ref{fig:example}~(a). 
This network has three attractors that are marked with dark grey nodes, including $A_1=\{000\}$,  $A_2=\{110\}$, and  $A_3=\{111\}$,  
The strong basin of each attractor is marked as the light grey region.
The weak basin of $A_1$ includes all the states except for states $110$ and $111$. 
The weak basin of $A_2$ and $A_3$ are 
$\bas_\ts^W(A_2)=\{010, 100, 101,110\}$
and $\bas_\ts^W(A_3)=\{011,101,111\}$. 
\end{example}

\section{Our Methods}
\label{sec:method}
\subsection{The control problem}
\label{ssec:problem}
As discussed in the introduction, 
direct cell reprogramming harnesses abundant somatic cells and reprograms them into desired deficient cells. 
However, a major obstacle to the application of this novel technique lies in the identification of effective targets,  
the intervention of which can lead to desired changes. 
We aim to solve this problem by identifying key molecules based on Boolean networks that model gene regulatory networks, 
such that the control of these molecules can drive the dynamics of a given network from a source attractor to the desired target attractor. 
We call it {\it source-target control} of Boolean networks.

Thanks to the rapid advances in gene editing techniques, 
the control can be applied for different periods of time.
Thus, we have {\it instantaneous control}, {\it temporary control} and {\it permanent control}, defined below.  
\begin{definition}[Instantaneous, temporary and permanent controls] \label{def:tempControl}

\noindent
(1) An instantaneous control is a control $C=(\zero, \one)$, 
such that by applying $C$ to $s$ instantaneously, 
the network always reaches the target attractor $\target$. \\
\noindent
(2) A temporary control is a control $C=(\zero, \one)$, such that there exists a $t_0 \geq 0$, 
for all $t \geq t_0$, the network always reaches the target attractor $\target$ on 
the application of $C$ to $s$ for $t$ steps.\\
\noindent
(3) A permanent control is a control $C=(\zero, \one)$, such that the network always reaches the target attractor $\target$ on the permanent application of $C$ to $s$.
\end{definition}

Temporary control applies perturbations for sufficient time and then is released,  
while permanent control maintains the perturbations for all the following time steps. 
Benefited from the extended intervention effects, 
temporary and permanent controls can potentially reduce the number of perturbations, 
which makes experiments easier to carry out and less costly~\cite{SPP19b}.

\begin{figure}[!t]
\subfloat[One-step control]{\label{fig:onestep}
\includegraphics[width=0.48\linewidth]{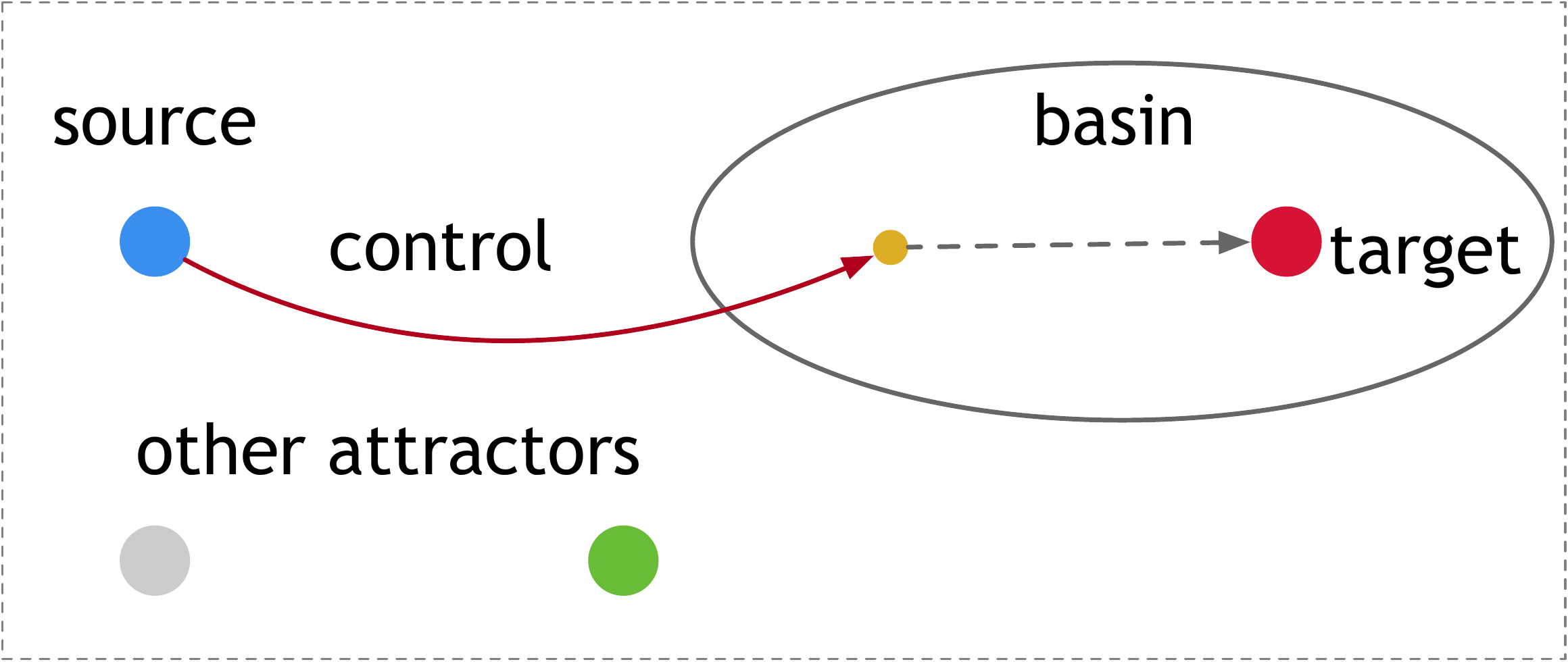} 
}
\subfloat[Sequential control]{\label{fig:seq}
\includegraphics[width=0.48\linewidth]{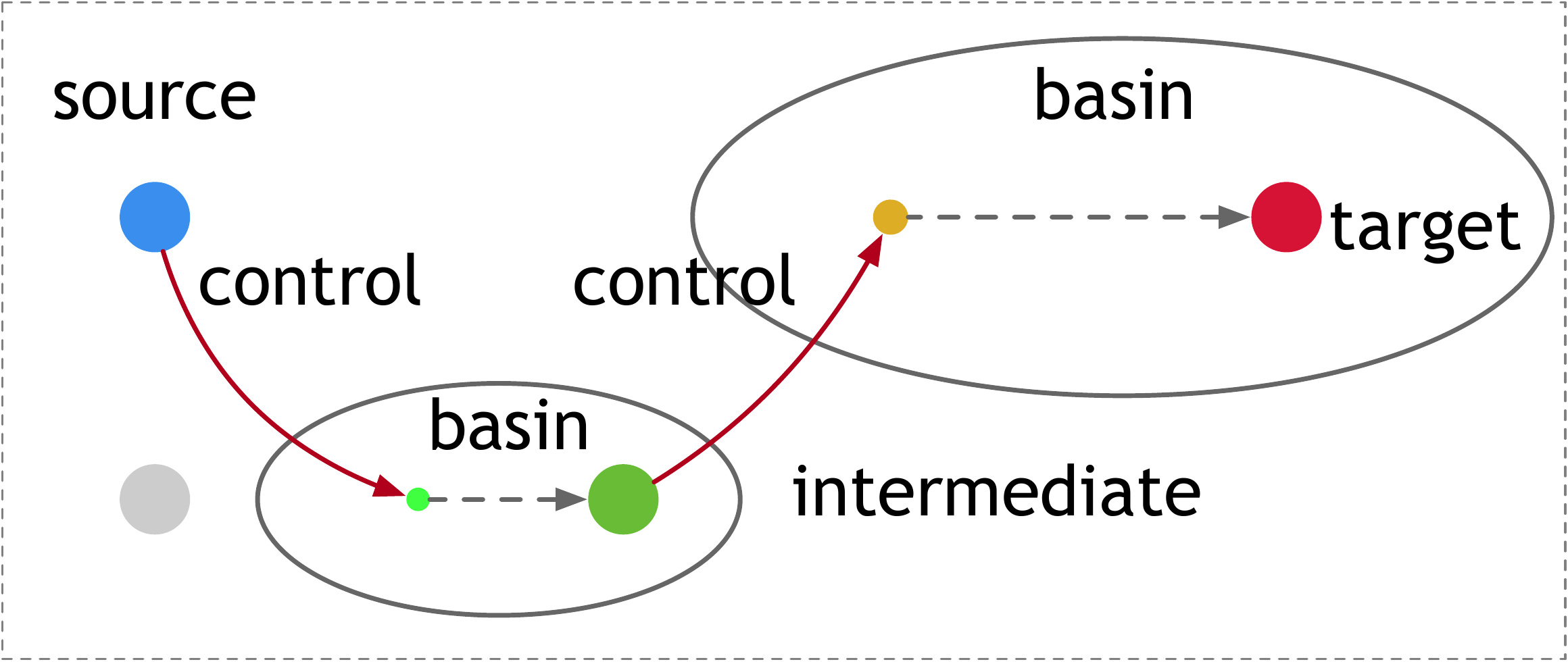}
}
\caption{Two control strategies.}
\label{fig:control_strategies}
\end{figure}

The source-target control can also be achieved in one step or in multiple steps, 
called {\it one-step control} and {\it sequential control}, respectively.  
As illustrated in Fig.~\ref{fig:onestep}, one-step control simultaneously applies all the required perturbations for one time (red arrow) to drive the network from a source state (blue node) to a state (yellow node), 
from which the network will converge spontaneously to the target attractor in finite time steps (dashed line). 
In Fig.~\ref{fig:seq},
sequential control utilises other states as intermediates and identifies a sequence of perturbations, 
the application of which guides the network towards the target attractor in a stepwise manner. 
Considering difficulties in conducting clinical experiments, 
we are interested in {\it attractor-based sequential control}, where only biologically observable attractors can act as intermediates.

Given a source attractor $\source$ and a target attractor $\target$ of $\ts$, 
the one-step control is formally defined as:
\begin{definition}[One-step control]
Compute a control $C_{A_s\rightarrow A_t}$, such that the application of $C_{A_s\rightarrow A_t}$ to a state $s\in \source$ can drive the network towards $\target$. 
\end{definition}

When the control $C_{A_s\rightarrow A_t}$ is the instantaneous, temporary or permanent control, 
we call it one-step instantaneous, temporary or permanent control (OI, OT or OP), respectively. 
To minimise the experimental cost, we are interested in the minimal solution $C^{\min}_{A_s\rightarrow A_t}$, where $C^{\min}_{A_s\rightarrow A_t}$ is the minimal such subset of $[n]$. 
Let $\mathcal{A}$ be the attractors of $\ts$. 
The attractor-based sequential control is defined as:

\begin{definition}[Attractor-based sequential control]
Find a sequence of attractors of $\ts$, i.e. $\{A_1, A_2,\ldots,A_m\}$,
where $A_1= \source, A_m=\target$, $A_i \neq A_j$ for any $i,j\in[1,m]$ and $2 \leq m \leq |\mathcal{A}|$, 
such that after the application of a sequence of minimal one-step controls 
$\{C^{\min}_{A_1\rightarrow A_2},C^{\min}_{A_2\rightarrow A_3},\ldots,C^{\min}_{A_{m-1}\rightarrow A_m} \}$, 
the network always eventually reaches $A_m$, i.e. $\target$.  
We call it an attractor-based sequential temporary path, denoted as 
$$\pc: A_1 \xrightarrow{C^{\min}_{A_ 1\rightarrow A_2}} A_2 \xrightarrow{C^{\min}_{A_ 2\rightarrow A_3}} A_3 \xrightarrow{\ldots} \ldots \xrightarrow{C^{\min}_{A_{m-1}\rightarrow A_m}} A_m$$
$(|C^{\min}_{A_1\rightarrow A_2}|+|C^{\min}_{A_2\rightarrow A_3}|+\ldots+|C^{\min}_{A_{m-1}\rightarrow A_m}|)$ is the total number of perturbations.
\end{definition}
Similarly, when the control $C_{A_s\rightarrow A_t}$ is the instantaneous, temporary or permanent control, 
we call it attractor-based sequential instantaneous, temporary or permanent control (ASI, AST or ASP), respectively.

We have developed efficient methods to tackle the minimal OI, OT and OP~\cite{PSPM18,PSPM19,SPP19b}, 
as well as ASI~\cite{MSHPP19,MSPPHP19}. 
Considering the advantages of sequential control and temporary and permanent perturbations,
in this paper we shall develop methods to solve the AST and ASP control problems.

\subsection{Attractor-based sequential temporary control}
\label{ssec:ast}
\begin{algorithm*}[!t]
\caption{Attractor-based sequential temporary control of BNs}
\label{alg:st}
\begin{algorithmic}[1]
\Procedure{Comp\_Seq\_Temp}{$F,k,\source,\target,\mathcal{A}$} 
\State Initialise a list $I := \emptyset$ to store possible intermediate attractors.
\State $\mathit{WB_\target} := ${\sc Comp\_Weak\_Basin}($F, \target$) \hfill{\it // weak basin of the target}
\State $\mathit{SB_\target} := ${\sc Comp\_Strong\_Basin}($F, \target$) \hfill{\it // strong basin of the target}
\State Initialise a dictionary to store paths $\mathcal{L} := \{L_{A_1}, L_{A_2}, \ldots, L_{A_m}\},~A_i\in \mathcal{A}$.

\For{$A \in (\mathcal{A}\setminus A_t)$} \hfill{\it //find attractors that have shorter paths to $\target$}
	\State $ C_{A \rightarrow \target} := ${\sc Comp\_Temp\_Control}$(A,\wb_\target,\sb_\target)$
	\If{($A = \source$ and $|C_{A \rightarrow \target}| \leq k$) or ($A \neq \source$ and $|C_{A \rightarrow \target}| \leq k-1$)} \\\hfill{\it // $C_{\source \rightarrow A}$ needs at least one perturbation}
		\State $\pa_{A\rightarrow \target}.$add$(\target)$ 
		\State $\pc_{A \rightarrow \target}.$add$(C_{A \rightarrow \target})$
		\State Add the path $(\pa_{A\rightarrow \target}, \pc_{A \rightarrow \target})$ to $L_A$	
		\State Add $A$ to $I$ as a candidate intermediate if $A \neq \source$. 
	\EndIf
\EndFor

\While{$I \neq \emptyset$}
	\State Initialise a new list $I' := \emptyset$
	\For{$A'_t \in I$} \hfill {\it // new target}
		\State $\mathit{WB_{A'_t}} := ${\sc Comp\_Weak\_basin}($F, A'_t$)
		\State $\mathit{SB_{A'_t}} := ${\sc Comp\_Strong\_basin}($F, A'_t$)
		
		\For{$A'_s  \in (\mathcal{A } \setminus (A'_t\cup A_t) )$} \hfill{\it // new source}
			\State $C_{A'_s \rightarrow A'_t} := ${\sc Comp\_Temp\_Control}$(A'_s ,\wb_{A'_t},\sb_{A'_t})$

			\For{$( \pa_{A'_t \rightarrow \target} ,\pc_{A'_t \rightarrow \target}) \in L_{A'_t}$ }
				\State $\pa_{A'_s\rightarrow A_t} := \pa_{A'_t \rightarrow \target}$; 
				Insert $A'_t$ to the beginning of $\pa_{A'_s\rightarrow A_t}$.

				\If{$A'_s \notin \pa_{A'_t\rightarrow A_t}$}
					\State Let $h$ denote the number of perturbations required by  $\pc_{A'_s\rightarrow A_t}$. 
					\If { ($A'_s=\source$ and $h\leq k$) or ($A'_s \neq \source$ and $h\leq k-1$) }
						\State $\pc_{A'_s \rightarrow A_t} := \pc_{A'_t\rightarrow \target}$ 
						\State Insert $C_{A'_s \rightarrow A'_t}$ to the beginning of $\pc_{A'_s \rightarrow A_t}$. 
						\State  Add the extended path $(\pa_{A'_s \rightarrow A_t}, \pc_{A'_t \rightarrow A_t})$ to $L_{A'_s}$.	
						\State Add $A'_s$ to $I'$ as a candidate intermediate if $A'_s \neq A_s$.
					\EndIf
				\EndIf
			\EndFor
		\EndFor
	\EndFor 
	\State $I := I'$
\EndWhile

\State Return $L_\source$
\EndProcedure

\Procedure{Perm\_Control\_Validation}{$C_{A'_s \rightarrow A'_t}, A'_t, \pa_{A'_t \rightarrow A_t}, \pc_{A'_t \rightarrow A_t}$}
	\State $A_1  :=  \pa[0]$ \hfill{\it // the first intermediate $A_1$ in $\pa_{A'_t \rightarrow A_t}$}
	\State $C_{A'_t \rightarrow A_1} := \pc[0]$ \hfill{\it // the first control set $C_{A'_t \rightarrow A_1}$ in $\pc_{A'_t \rightarrow A_t}$}
	\State $\pa' := \pa_{A'_t \rightarrow A_t}.$pop$()$, $\pc' := \pc_{A'_t \rightarrow A_t}.$pop$()$
	\hfill{\it //delete the first element}
	\State $C'' := C_{A'_s \rightarrow A'_t} \setminus C_{A'_t \rightarrow A_1}$
	\State $\flag := \True$
	\If{$A'_t|_{C''} = A_1|_{C''}$ and $\pa' \neq \emptyset$}
		\State $\flag := ${\sc Perm\_Control\_Validation}$(C'', A'_t, \pa', \pc')$
	\ElsIf{$A'_t|_{C''} \neq A'|_{C''}$}
		\State $\flag :=\False$
	\EndIf

	\Return $\flag$
\EndProcedure
\end{algorithmic}
\end{algorithm*}

Algorithm~\ref{alg:st} describes a procedure {\sc Comp\_Seq\_Temp} to compute AST control paths within $k$ perturbations. 
This algorithm is based on our previously proposed methods, 
including the computation of weak basin and strong basin~\cite{PSPM19,PSPM18}, denoted {\sc Comp\_Weak\_Basin} and {\sc Comp\_Strong\_Basin}, 
and the computation of minimal OT control~\cite{SPP19b}, namely {\sc Comp\_Temp\_Control}. 
Particularly, the procedure {\sc Comp\_Temp\_Control} is based on the following theorem.

\begin{theorem} \label{theo:tempControl}
A control $C=(\zero,\one)$ is a minimal temporary control from $s$ to $\target$ iff (1)
$\bas_\ts^S(\target) \cap S_C \neq \emptyset$ and $C(s) \in \bas_{\ts|_C}^S(\bas_\ts^S(\target)\cap S|_C)$ and (2) $C$ is a minimal such subset of $\{1,2,\ldots,n\}$.
\end{theorem}

The procedure {\sc Comp\_Seq\_Temp} takes as inputs the Boolean functions $F$, a threshold $k$ of the number of perturbations, a source attractor $\source$, a target attractor $\target$, and the set of attractors $\mathcal{A}$ of $\ts$.
It contains two parts.

The first part includes lines $2$-$13$. 
We compute the minimal OT control set from attractor $A$ $(A\in \mathcal{A}$ and $A\neq \target)$ to $\target$. 
For each attractor $A$, we generate a dictionary $L_A$ to save all the valid sequential control paths from $A$ to $\target$ (line $5$ and $12$). 
The OT control $C_{A \rightarrow \target}$ from $A$ to $\target$ is considered valid and saved to $L_A$ if 
(1) $A$ is the source attractor $\source$ and the number of perturbations $|C_{A \rightarrow \target}|$ is not greater than $k$; 
or (2) $A$ is not $\source$ and $|C_{A \rightarrow \target}|$ is less or equal to $(k-1)$. 
If $A$ is an intermediate attractor ($A\neq \source$), 
$C_{\source \rightarrow A}$ requires at least one perturbation. 
Therefore, the size of $C_{A \rightarrow \target}$ should not exceed $(k-1)$. 
$A$ is saved to $I$ as an intermediate attractor if $A \neq \source$ and $|C_{A \rightarrow \target}| \leq k-1$.

The second part includes lines $14$-$30$. 
We extend the control paths computed in the previous part by recursively taking every intermediate attractors $A'_t \in I$ as a new target and computing the minimal temporary control from an attractor $A'_s~(A'_s\in (\mathcal{A} \setminus (A'_t\cup \target) ))$ to $A'_t$. 
Specifically, for each new target attractor $A'_t$, 
we compute the minimal temporary control set $C_{A'_s \rightarrow A'_t}$ from $A'_s$ to $A'_t$ (line $20$). 
Then, for every sequential path from $A'_t$ to $A_t$, for instance $(\pa_{A'_t \rightarrow A_t}, \pc_{A'_t \rightarrow A_t})$, 
we verify whether $A'_s$ can be appended to the beginning of $\pa_{A'_t \rightarrow A_t}$ to form a new path from $A'_s$ to $A'_t$
based on the following two conditions: 
(1) $A'_s$ is not an intermediate in path $A'_t\rightarrow \ldots \rightarrow A_t$; and   
(2) the total number of perturbations of the new path $\pa_{A'_s \rightarrow A_t}$ should not exceed $k$ (or $k-1$) 
if $A'_s=\source$ (or $A'_s \neq \source$). 
When both conditions are satisfied, we save the new path to $L_{A'_s}$ (line $28$) and add $A'_s$ to $I'$ as a new candidate intermediate if $A'_s \neq \source$ (line $29$). 
After going through all the intermediate attractors in $I$ (lines $16-29$), 
we update the set of intermediate attractors $I$ and repeat steps at lines $14$-$30$ until $I$ is an empty set.

\subsection{Attractor-based sequential permanent control}
\label{ssec:asp}
In this section, we develop an algorithm to solve the ASP control problem. 
We have developed an algorithm to compute the minimal OP control~\cite{SPP19b}, 
denoted as {\sc Comp\_Perm\_Control}, based on the following theorem. 
\begin{theorem} \label{theo:permControl}
A control $C=(\zero,\one)$ is a minimal  permanent control from $s$ to $\target$ iff (1) $C(s) \in \bas_{\ts|_C}^S(\target)$ and (2) $C$ is a minimal such subset of $\{1,2,\ldots,n\}$. 
\end{theorem}

The algorithm for ASP control explores the same way as Algorithm~\ref{alg:st} to construct sequential paths, but it is more involved. 
It can be achieved by modifying procedure {\sc Comp\_Seq\_Temp} in Algorithm~\ref{alg:st} as follows. 
First, at lines $7$ and $20$, we simply replace the procedure {\sc Comp\_Temp\_Control} with the procedure {\sc Comp\_Perm\_Control}. 
Second, when extending the sequential paths, besides the conditions at line $25$, 
we add the procedure {\sc Perm\_Control\_Validation} in Algorithm~\ref{alg:st} 
 to verify whether the control $C_{A'_s \rightarrow A'_t}$ can be inserted to the beginning of $\pc_{A'_t \rightarrow \target}$. 
Because for each control step of AST, the temporary perturbations are released at one time point to retrieve the original transition system and let the network evolve spontaneously to the the intermediate/target attractor.  
But ASP adopts permanent control that will be maintained for all the following time steps. 
Therefore, when extending a permanent control $C$ to the beginning of a sequential path, 
it has to be verified whether the application of $C$ will affect the reachability of the following control steps. 
To avoid duplication, here we only give the explanations of the procedure {\sc Perm\_Control\_Validation}. 
The purpose of this procedure is to verify whether the control $C_{A'_s \rightarrow A'_t}$ can be added to the beginning of $\pa_{A'_t \rightarrow A_t}$
to form a new path $\pa_{A'_s \rightarrow A_t}$
The verification is carried out recursively. 
Let us assume $\pa_{A'_t \rightarrow A_t}=\{A_1,A_2,\ldots,\target\}$. 
The first intermediate attractor is $A_1$ and the control from $A'_t$ to $A_1$ is $C_{A'_t \rightarrow A_1}$. 
Since $C_{A'_s \rightarrow A'_t}$ and $ C_{A'_t \rightarrow A_1}$ may require to perturb the same node in the opposite way, 
we compute $C_{A'_s \rightarrow A'_t}$ set minus $ C_{A'_t \rightarrow A_1}$ and denote it as $C''$. 
If the projections of $A'_t$ and $A_1$ to $C''$ are the same, $A_1$ is preserved under the permanent control $C''$ and we proceed to the remaining control steps (lines $38$-$39$); 
otherwise, $C_{A'_s \rightarrow A'_t}$ is not a valid sequential permanent control (lines $40$-$41$).

\begin{example}
\label{eg:bn2}
To continue with Example~\ref{eg:bn}, we compute the control paths from $A_1$ to $A_3$ with ASI, AST and ASP control methods.  
For this case, AST and ASP have the same results. 
The shortest ASI control path is
$ A_1 \xrightarrow{\{x_1,x_2\}} A_2 \xrightarrow{\{x_3\}} A_3$, which needs three perturbations.   
There are two shortest AST/ASP control paths: $ A_1 \xrightarrow[\{x_1\}]{\{x_2\}} A_2 \xrightarrow{\{x_3\}} A_3$, which require two perturbations in total. 
\end{example}

\section{Evaluation}
\label{sec:evaluation}
In this section, we evaluate the performance of AST and ASP
on several real-life biological networks. 
To demonstrate their efficacy, 
we compare their performance with ASI~\cite{MSHPP19}. 
The minimal number of perturbations required by OI, OT and OP is set as the threshold $k$ of the number of perturbations for ASI, AST and ASP, respectively. 
In this way, the results will demonstrate whether AST and ASP can find sequential paths with fewer perturbations than ASI. 
All the methods are implemented as an extension of our software tool ASSA-PBN~\cite{MPSY18} 
and all the experiments are performed on a high-performance computing (HPC) platform, 
which contains CPUs of Intel Xeon Gold 6132 @2.6 GHz. 
We describe and discuss the results of the myeloid differentiation network~\cite{KMST11}
and the Th cell differentiation network~\cite{NCCT10} in detail (Sections~\ref{ssec:myeloid} and~\ref{ssec:thcell}),
and we give an overview of the results of the other networks (Section~\ref{ssec:others}).

\subsection{The myeloid differentiation network} 
\label{ssec:myeloid}
The myeloid differentiation network is constructed to model the differentiation process of common myeloid progenitors (CMPs) into four types of mature blood cells~\cite{KMST11}.
With our attractor detection method~\cite{MPQY19},  
we identify six single-state attractors of the network,
five of which are non-zero attractors (not all the nodes have a value of `0'). 
It has been validated that expressions of four attractors correspond to microarray expression profiles of megakaryocytes, erythrocytes, granulocytes and monocytes~\cite{KMST11}.
The fifth attractor with the activation of PU1, cJun and EgrNab might be caused by pathological alterations~\cite{KMST11} 
and the sixth attractor is an all-zero attractor, where all the nodes have a value of `0'.  

\begin{figure}[!t]
\subfloat[ASI]{\label{fig:myeloid-asi}
\includegraphics[width=0.45\linewidth]{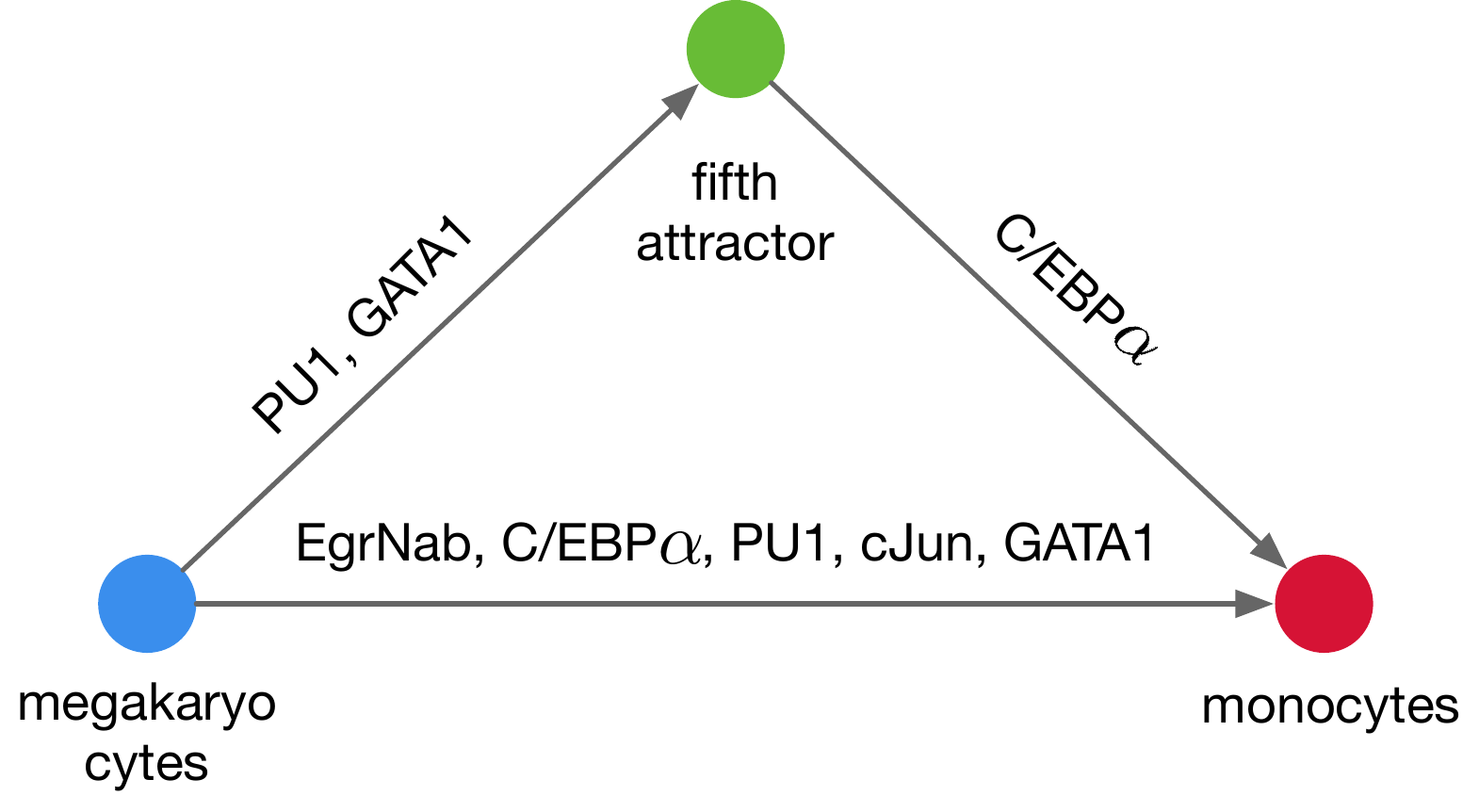} 
}
\subfloat[AST/ASP]{\label{fig:myeloid-astp}
\includegraphics[width=0.45\linewidth]{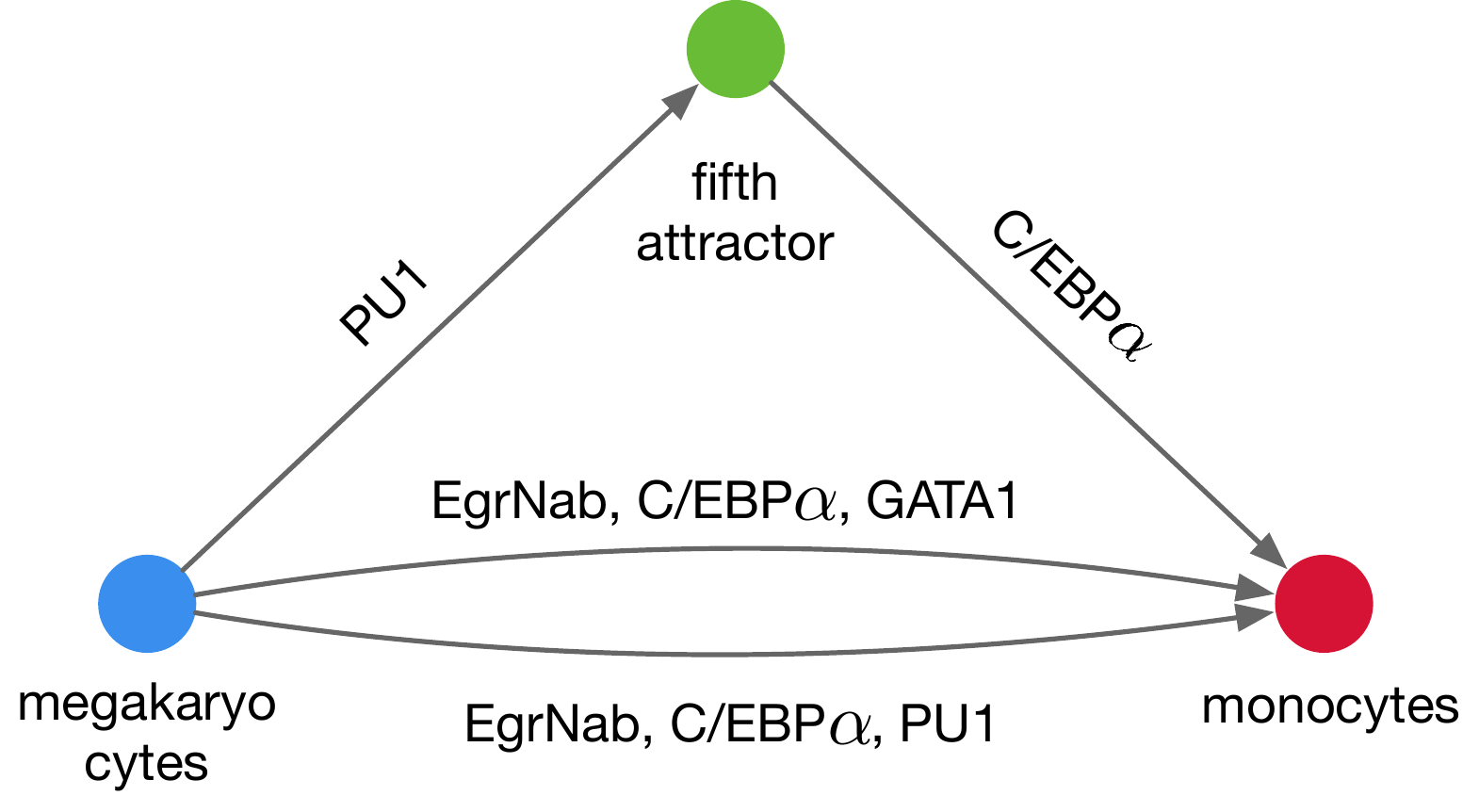}
}
\caption{Control of the myeloid differentiation network.}
\label{fig:myeloid}
\end{figure}

We take the conversion from megakaryocytes to granulocytes as an example to show the performance of the methods. 
Note that the sixth attractor does not have a biological interpretation 
and mature erythrocytes in mammals do not have cell nucleus, therefore  
we do not consider these two attractors as intermediate attractors. 
Under this condition, the three methods (ASI, AST, ASP) identify both one-step and sequential paths as illustrated in Fig.~\ref{fig:myeloid}.    
In particular, the results of AST and ASP are identical.  
We can see that the minimal OI control  
requires the activation of EgrNab, C/EBP$\alpha$, PU1, cJun and the inhibition of GATA1 (Fig.~\ref{fig:myeloid-asi}); 
while OT or OP can achieve the goal by either (1) the activation of EgrNab, C/EBP$\alpha$ and PU1; 
or (2) the activation of EgrNab and C/EBP$\alpha$, together with the inhibition of GATA1 (Fig.~\ref{fig:myeloid-astp}). 
All the sequential paths need two steps, where the fifth attractor is adopted as an intermediate attractor. 
For the first step, ASI activates PU1 and inhibits GATA1, 
while AST or ASP only needs to activate PU1. 
When the network converges to the fifth attractor, 
all the three methods require to activate C/EBP$\alpha$. 
After that, the network will evolve spontaneously to the target attractor monocytes. 
Fig.~\ref{fig:myeloid} shows that AST and ASP are able to 
identify a path with only two perturbations, 
while ASI requires at least three perturbations.   

The efficacy of the identified sequential temporary/permanent path is confirmed by the predictions in~\cite{KMST11}. 
According to the expression profiles, both PU1 and C/EBP$\alpha$ are not expressed in MegE lineage (megakaryocytes and erythrocytes), 
while they are expressed in GM lineage (monocytes and granulocytes). 
In this network, 
no regulator can activate C/EBP$\alpha$ and PU1 is primarily activated by C/EBP$\alpha$. 
Therefore, C/EBP$\alpha$ has to be altered externally to reprogram MegE lineage to GM lineage. 
However, more perturbations are necessary to accurately reach the monocytes lineage. 
Sustained activation of PU1 and the absence of C/EBP$\alpha$ guide the network to the fifth attractor, 
the expression of which differs with monocytes only in C/EBP$\alpha$~\cite{KMST11}. 

\subsection{The Th Cell differentiation network} 
\label{ssec:thcell}
The T-helper (Th) cell differentiation network is a comprehensive model integrating regulatory network and signalling pathways that regulate Th cell differentiation~\cite{NCCT10}.  
This network consists of 12 single-state attractors under one initial condition  
and the attractors can be classified into different Th subtypes based on the expression of four master regulators (TBET, GATA3, PORGT and FOXP3)~\cite{NCCT10}. 

\begin{figure}[!t]
\centering
\scalebox{.9}{
\subfloat[ASI]{\label{fig:T-diff-asi}
\includegraphics[width=0.31\linewidth]{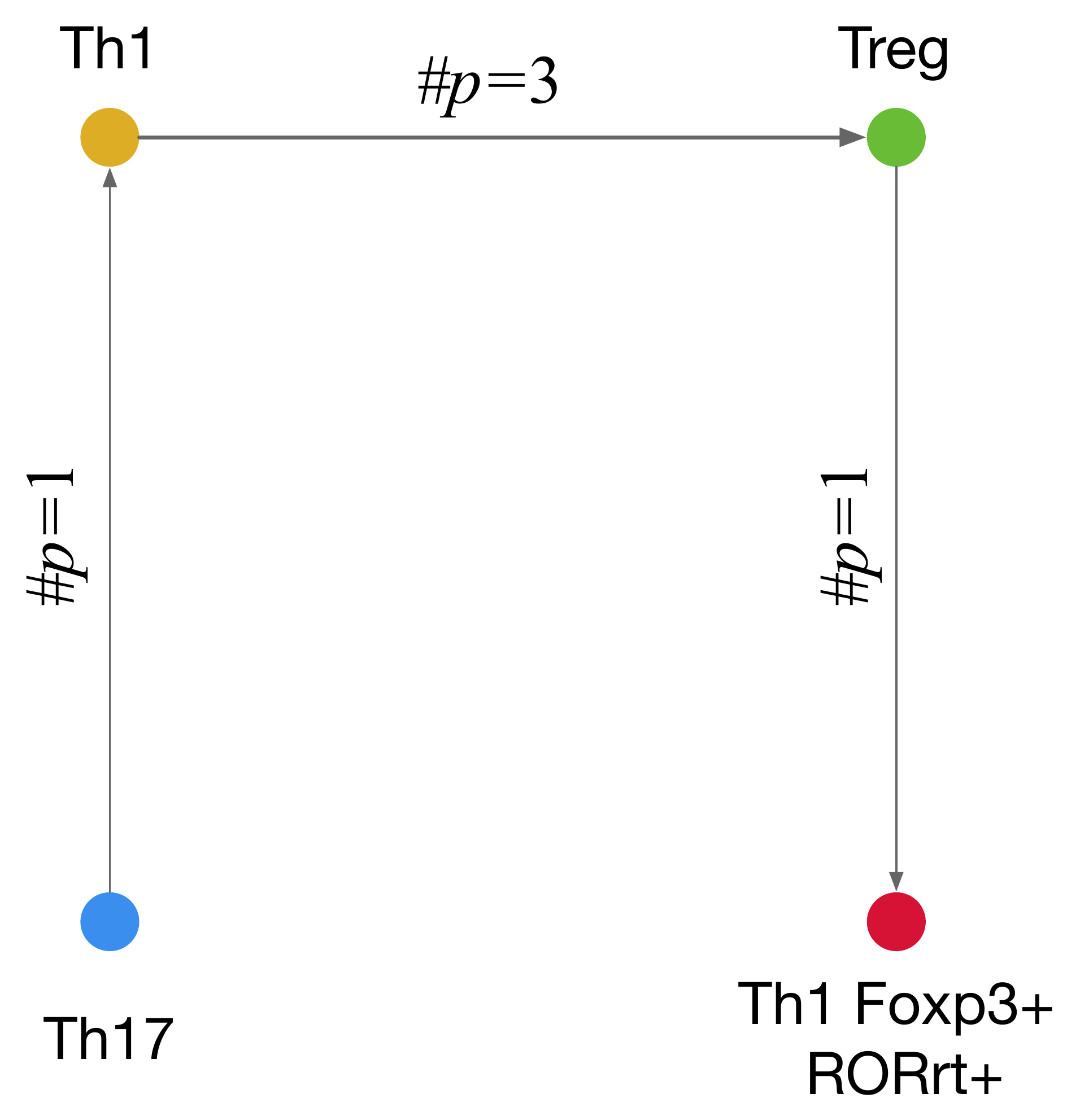} 
}
\subfloat[AST]{\label{fig:T-diff-ast}
\includegraphics[width=0.31\linewidth]{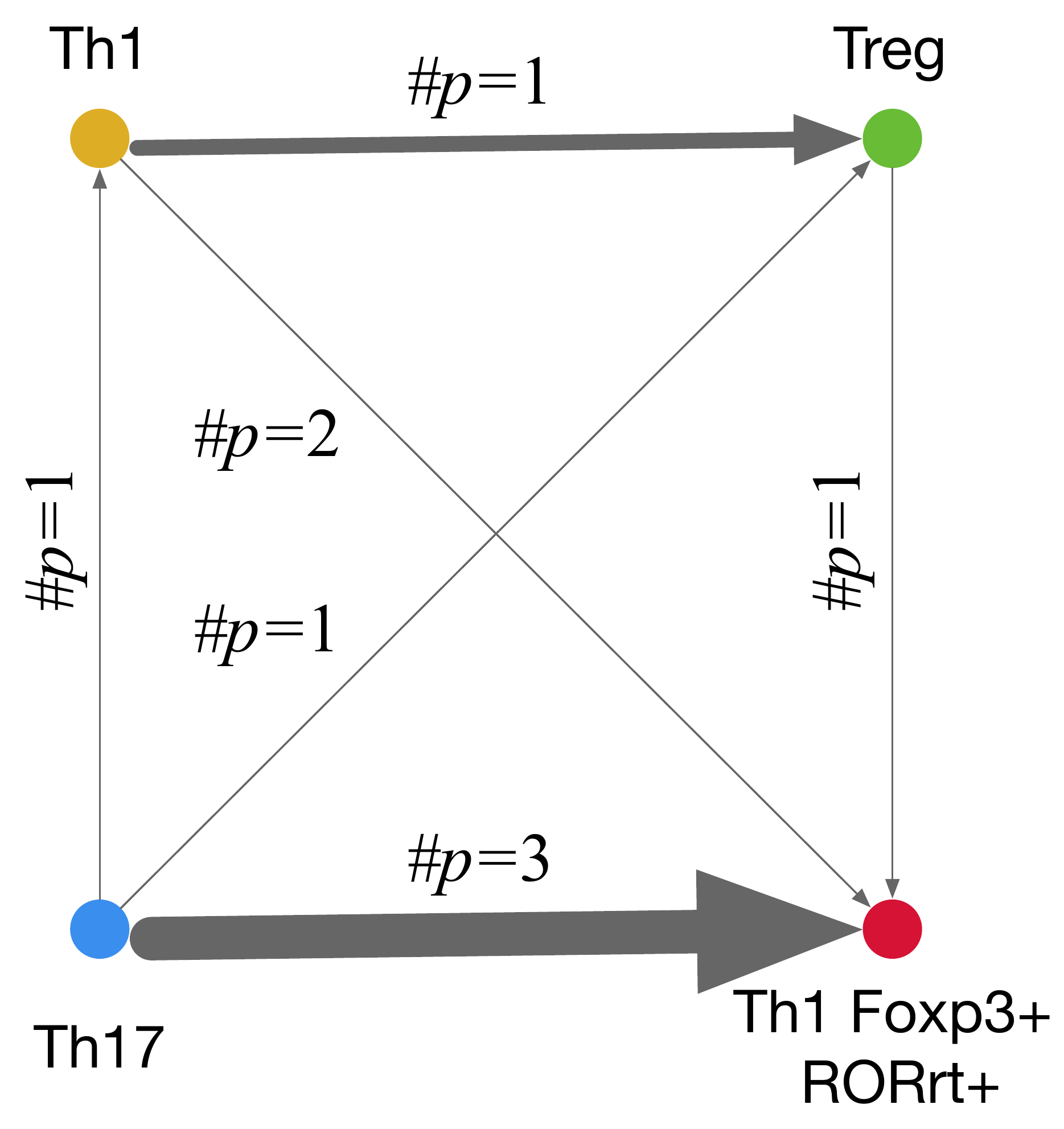}
}
\subfloat[ASP]{\label{fig:T-diff-asp}
\includegraphics[width=0.31\linewidth]{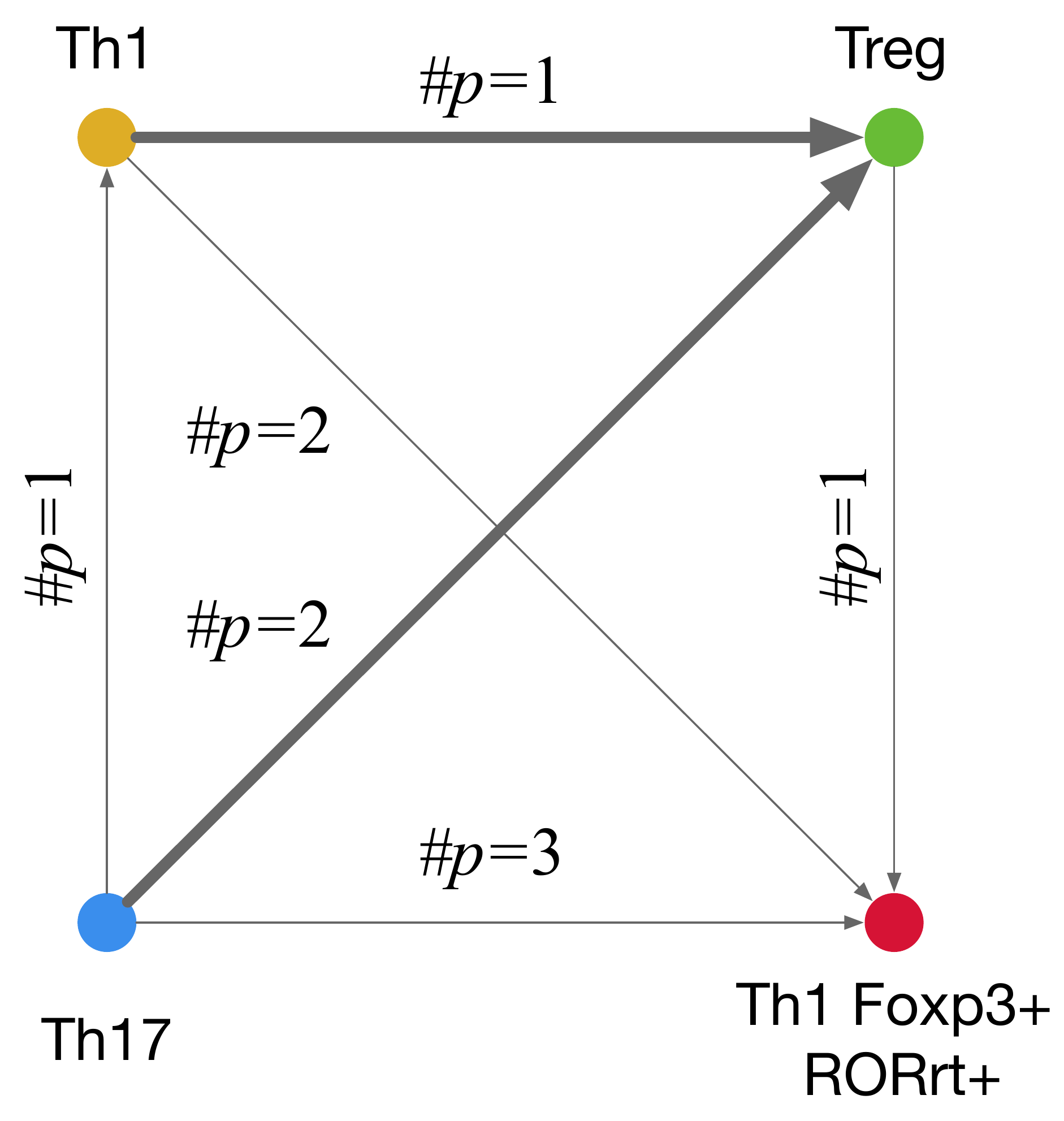} 
}
}
\caption{Control of the Th cell differentiation network.}
\label{fig:T-diff}
\end{figure}

Let Th17 and a Th1 subtype (Th1 Foxp3+ RORrt+) be the source and target attractors, respectively.
For the purpose of illustration, we limit the number of control paths by only adopting Th1 and Treg as intermediate attractors. 
In addition, we set the node `proliferation' as a non-perturbed node, 
since it denotes a cell fate and thus cannot be perturbed in reality.
Fig.~\ref{fig:T-diff} describes the control paths identified by the three methods.   
The thickness of arrows implies the number of control sets and 
the equations $\#p=m$ above each arrow denotes the number of perturbations required by each step.  
All the methods identify sequential paths passing through Th1 and/or Treg. 
Fig.~\ref{fig:T-diff-asi} only shows the shortest ASI path with five perturbations, 
while AST and ASP provide multiple paths with only two or three perturbations (Fig.~\ref{fig:T-diff-ast} and Fig.~\ref{fig:T-diff-asp}), 
demonstrating the advantages of AST and ASP in reducing the number of perturbations.  
Among the sequential paths of AST and ASP, only the AST path, $\text{Th17}~\xrightarrow{\text{IL27R}}\text{Treg} \xrightarrow{\text{TBET}} \text {Th1 subtype}$, perturbs two nodes, 
all the other paths using either temporary or permanent perturbations need to perturb at least three nodes. 
This shows that AST has the potential to further reduce the number of perturbations compared to ASP. 
Moreover, in terms of the number of solutions, 
it is obvious that the arrows in Figure~\ref{fig:T-diff}(b) are thicker than those in Figure~\ref{fig:T-diff}(c), which indicates that AST provides more solutions than ASP.

\subsection{Other biological networks} 
\label{ssec:others}
Besides the myeloid and Th cell differentiation networks, 
we also apply the three control methods to several other biological networks~\cite{HGZKK12,SFLK09,OKB16,COOAD17,RRC15,COAM15,CHSLL14}. 
Here is a brief introduction of the networks. 
\begin{itemize}
	\item The cardiac gene regulatory network integrates key regulatory factors that play key roles in early cardiac development and FHF/SHF determination~\cite{HGZKK12}. 
	\item The ERBB receptor-regulated G1/S transition network is built to identify efficacious targets for treating trastuzumab resistant breast cancer~\cite{SFLK09}. 
	\item The network of PC12 cell differentiation is built to capture the complex interplay of molecular factors in the decision of PC12 cell differentiation~\cite{OKB16}. 
	\item The network of hematopoietic cell specification is constructed to capture the lymphoid and myeloid cell development~\cite{COOAD17}.
	\item The network of bladder tumour is constructed to study mutually exclusivity and co-occurrence in genetic alterations~\cite{RRC15}. 
	\item The pharmacodynamic model of bortezomib responses integrates major survival and apoptotic pathways in U266 cells to connect bortezomib exposure to multiple myeloma cellular proliferation~\cite{COAM15}.
	\item The network of a CD$4^+$ immune effector T cell is constructed to capture cellular dynamics and molecular signalling under both immunocompromised and healthy settings~\cite{CHSLL14}. 
\end{itemize}
Columns $2$-$4$ of Table~\ref{tab:othernetworks} summarise the number of nodes, edges and attractors contained in each network.
For each network, we choose a pair of source and target attractors and compute control paths with ASI, AST and ASP. 

\smallskip\noindent
\textbf{Efficacy.} 
For each pair of source and target attractors, all the control paths with at most $k$ perturbation are computed. 
For the purpose of comparison, in Table~\ref{tab:othernetworks}, 
columns $5$-$7$ only summarise the minimal number of perturbations needed by each control method 
and columns $8$-$10$ summarise the number of corresponding control paths. 
It shows that by extending the period of control time, 
AST and ASP have the ability to compute more control paths with fewer perturbations than ASI. 
This brings significant benefits for practical applications.   
First, fewer perturbations can reduce the experimental costs and make the experiments easier to conduct.
Second, a richer set of control paths provides biologists more options to tackle diverse biological systems. 

To further compare AST and ASP, AST is more appealing than ASP. 
As discussed in the previous subsection, the control of Th cell differentiation network shows that AST has the potential to identify smaller control sets than ASP. 
For the other cases listed in Table~\ref{tab:othernetworks}, although AST requires the same number of perturbations as ASP, 
AST identifies more solutions than ASP. 
Apart from that, AST has an intrinsic advantage compared to ASP --  
temporary control will eventually be released and therefore can eliminate risks of unforeseen consequences, 
which may be caused by the permanent shift of the dynamics. 

\smallskip\noindent
\textbf{Efficiency.} 
The last three columns of Table~\ref{tab:othernetworks} give the computation time of ASI, AST and ASP. 
Although AST and ASP take longer time than ASI, 
they are still quite efficient and are capable of handling large-scale and comprehensive networks. 
In general, the computational time of the methods depends on the size of the network, 
the threshold of the number of perturbations $k$ and the number of existing solutions within the threshold. 
By increasing the threshold $k$, our methods can identify more candidate solutions at the cost of longer computational time. 
Currently, due to the lack of large and well-behaved networks, we are not yet able to find out the precise limit of our methods on the size of networks.

\begin{table}[!t]
\centering
\begin{tabular}{|L{1.5cm}|R{0.6cm}|R{0.6cm}|R{0.6cm}|R{0.6cm}R{0.6cm}R{0.5cm}|R{0.6cm}R{0.6cm}R{0.6cm}|R{1.2cm}|R{1.2cm}|R{1.2cm}|}
\hline
\multirow{2}{*}{network} & \multirow{2}{*}{$|V|$} & \multirow{2}{*}{$|E|$}  
&\multirow{2}{*}{$|\mathcal{A}|$} & \multicolumn{3}{|c|}{\footnotesize \#perturbations} & \multicolumn{3}{|c|}{\# paths} & \multicolumn{3}{|c|}{time (seconds)} \\ \cline{5-13}
&&&& ASI & AST & ASP & ASI & AST & ASP& ASI & AST & ASP\\ \hline 
myeloid & $11$ & $30$ & $6$ & $3$ & $2$ & $2$ & $1$ & $1$ & $1$ & $0.006$ & $0.034$ & $0.038$ \\
cardiac & $15$ & $39$ & $6$ & $3$ & $2$ & $2$ & $1$ & $3$ & $2$ & $0.018$ & $0.658$ & $0.653$ \\ 
ERBB & $20$ & $52$ & $3$ & $8$ & $3$ & $3$ & $2$ & $3$ & $3$ & $0.007$ & $0.249$ & $0.319$ \\ 
PC12 & $33$ & $62$ & $7$ & $8$ & $2$ & $2$ & $3$ & $50$ & $30$ & $0.050$ & $1.188$ & $1.462$ \\ 
HSC & $33$ & $88$ & $5$ & $12$ & $2$ & $2$ & $2$ & $12$ & $6$ & $0.406$ & $12.217$ & $8.879$ \\ 
bladder & $35$ & $116$ & $4$ & $5$ & $2$ & $2$ & $2$ & $2$ & $2$ & $0.139$ & $0.709$ & $0.676$ \\ 
bortezomib & $67$ & $135$ & $5$ & $3$ & $2$ & $2$ & $1$ & $4$ & $2$ & $1.900$ & $105.184$ & $119.138$ \\
T-diff & $68$ & $175$ & $12$ & $5$ & $2$ & $3$ & $4$ & $1$ & $14$ & $9.713$ & $95.211$ & $71.044$ \\   
CD$4^+$ & $188$ & $380$ & $6$ & $3$ & $2$ & $2$ & $3$ & $48$ & $6$ & $256.492$ & $539.868$ & $1304.490$ \\ 

\hline
\end{tabular}
\caption{Control of several biological networks.}
\label{tab:othernetworks}
\end{table}

\section{Discussion}
\label{sec:discussion}
We have demonstrated the potential strengths of AST and ASP, however,  
they are not warranted to be the best methods for all kinds of biological systems. 
Indeed, there is no control method that can perfectly solve all the control problems 
due to the intrinsic diversity and complexity of biological systems. 
Given a specific task, it is thus recommended to compute all the control paths 
with available control methods. 
Various sets of identified therapeutic targets serve as candidates, 
such that biologists can choose appropriate targets, the modulation of which will not disrupt physiological functions of biological systems. 

Although the dynamics of asynchronous BNs are non-deterministic, 
our methods guarantee to find the shortest control paths with $100\%$ reachability {\it in silico}.  
Experimental validation is necessary to verify their therapeutic efficacy {\it in vivo}. 
It is worth noticing that the consistency of the efficacy {\it in silico} and {\it in vivo} highly relies on the quality of the constructed BNs. 
The identified perturbations can effectively modulate the dynamics as expected, 
provided that the adopted network well captures the structural and dynamical properties of the real-life biological system. 
However, mathematical modelling of vastly complex biological systems is already a challenging task by itself in systems biology. 
We have spotted some flaws of the constructed networks in the literature during analysis, summarised as follows.

First, simulation is often used to evaluate the stable behaviour of dynamics in most of the works.
However, simulation can hardly cover the entire transition system of a BN, which is exponential in the size of the network. 
As a consequence, the information on attractors is usually incomplete, especially for networks of medium or large sizes. 
This problem can be solved by using our attractor detection method~\cite{MPQY19,YMPQ19b} to identify all the exact attractors of a network.

Second, we noticed that the attractors of some large constructed networks are purely induced by input nodes. 
For instance, given a network with 2 input nodes (nodes without upstream regulators), it has $2^2$ attractors. 
Each attractor corresponds to one combination of the input nodes ($00,01,10,11$).
For such networks, the input nodes, that have different values in the source and target attractors, 
are the key nodes for modulating the dynamics. 
Such kind of networks may capture some activation or inhibition regulations, 
but they fail to depict the intrinsic mechanisms of biological processes.

Third, in some networks, cell phenotypes or cell fates, such as apoptosis, proliferation, and differentiation, 
are represented as marker nodes. 
Benefited from this, attractors can be classified based on the expressions of those nodes. 
However, a problem that often occurs is that there does not exist any control sets without perturbing these marker nodes. 
Again, we hypothesise that these constructed networks do not reflect the intrinsic properties of biological systems.

Our methods~\cite{MPQY19,YMPQ19b,PSPM19,SPP19b,MSHPP19} can provide accurate information of the networks,
such as the number and size of the attractors and potential sets of control nodes.  
Such information related to the network dynsmics should be taken into account
when inferring the networks by updating the Boolean functions or adding/deleting regulators.

\section{Conclusion and Future Work}
\label{sec:conclusion}

In this work, we have developed the attractor-based sequential temporary and permanent control methods to identify the shortest sequential control paths for modulating the dynamics of biological systems. 
To make it practical, only biologically observable attractors are served as intermediates. 
We compared the performance of the two methods with the attractor-based instantaneous control on a variety of biological networks.  
The results show that these two methods have apparent advantages in reducing the number of perturbations and enriching the diversity of solutions.  

Until now, we have developed source-target control methods to alter the dynamics of BNs in different ways. 
Currently, we are working on a target control method to identify a subset of nodes, 
the intervention of which can transform any somatic cells to the desired cell type. 
We also plan to study the control of probabilistic Boolean networks~\cite{SD10,TMPTSS13} based on our control methods for BNs. 
We believe our works can provide deep insights into regulatory mechanisms of biological processes 
and facilitate direct cell reprogramming.

\bibliographystyle{splncs04}
\bibliography{control}

\end{document}